\documentstyle[preprint,eqsecnum,aps,psfig]{revtex}
\begin{document}
\draft

\title{Ising spins coupled to a four-dimensional discrete Regge skeleton}
\author{E.~Bittner}
\address{Atominstitut der \"Osterreichischen Universit\"aten,
 TU Wien, A-1040 Vienna, Austria\\
Institut f\"ur Theoretische Physik, Universit\"at Leipzig, D-04109 Leipzig,
Germany}
\author{W.~Janke}
\address{ Institut f\"ur Theoretische Physik, Universit\"at Leipzig, D-04109 Leipzig,
Germany}
\author{H.~Markum }
\address{Atominstitut der \"Osterreichischen Universit\"aten,
 TU Wien, A-1040 Vienna, Austria}

\date{\today}
\maketitle
\begin{abstract}
Regge calculus is a powerful method to approximate a continuous manifold by 
a simplicial lattice,
keeping the connectivities of the underlying lattice fixed and
taking the edge lengths as degrees of freedom.
The discrete Regge model employed in this work limits the choice of the link lengths to
a finite number. To get more precise
insight into the behavior of the four-dimensional discrete Regge model,
we coupled spins to the fluctuating manifolds.
We examined the phase transition of the spin system and the associated
critical exponents. The results are obtained from finite-size scaling
analyses of Monte Carlo simulations.
We find consistency
with the mean-field theory of the Ising model on a static four-dimensional
lattice.

\end{abstract}

%
%
%

\narrowtext

\section{Introduction} \label{intro}

Spin systems coupled to fluctuating manifolds are analyzed as a simple example for matter
fields coupled to Euclidean quantum gravity. 
The gravitational action is unbounded from below due to conformal fluctuations. But
that does not necessarily render its quantum theory useless or the path integral undefined.
Indeed, the existence of a well defined phase was the first and probably most important result
of the numerical simulations in four dimensions~\cite{hamber84,berg}.
Its existence and
stability were explored in some detail using the standard Regge calculus
with continuous link lengths. It turned out that the well defined
phase is stable against variations of the measure and the lattice size~\cite{beirl94}.

In the discrete Regge model the problem of an unbounded action is not present as in
standard Regge calculus. Because of the restriction of possible quadratic link lengths to
two values~\cite{beirlz2,fleming} in the discrete Regge model the action can only reach an extreme but finite value. 
The expectation values do not diverge if the well defined phase is left. What happens is that
the lattice ``freezes'' at large positive and negative values of the gravitational coupling,
as expected for a spin system~\cite{homo}. To get more precise
ideas about the behavior of the four-dimensional discrete Regge model, we coupled Ising spins to the fluctuating manifolds.
We examined the phase transition of the spin system and the associated
critical exponents.

The rest of the paper is organized as follows. In Sec.~\ref{sect2} we introduce the discrete
Regge model and give some details of the analyzed observables. The results of the Monte Carlo
simulations  are presented in Sec.~\ref{sect3}, and concluding remarks can be found in 
Sec.~\ref{sect4}.

\section{Models and observables} \label{sect2}

The situation for the discrete Regge model is both structurally and computationally much
simpler than the standard Regge calculus with continuous link lengths.
The restriction of the edge lengths to just two values was carefully
examined in 2D where an interpolation from $Z_2$ to $Z_\infty$ was
performed~\cite{bittnerzn}. It turned out that the phase transition with
respect to the cosmological constant is universal. This was tested for pure
gravity in 2D and is expected to be the case also in 4D. Compared with 
standard Regge calculus, numerical simulations of the discrete Regge model
can be done more efficiently by implementing look-up
tables and using the heat-bath algorithm. In the actual computations we took
the squared link lengths as
$q_{ij} \equiv q_l = b_l ( 1+\epsilon \sigma_l)$ with $\sigma_l \pm 1$.
The Euclidean triangle inequalities are satisfied automatically as long as 
$\epsilon < \epsilon_{\rm max}$.
Because a four-dimensional Regge
skeleton with equilateral simplices cannot be embedded in flat space,
$b_l$ takes different values depending on the type of the edge $l$. In particular
$b_l=1,2,3, 4$ for edges, face diagonals, body diagonals, and the hyperbody
diagonal of a hypercube.

We investigated the partition function
\begin{equation}\label{z}
Z=\sum_{\{s\}}\int D[q] \exp\left[-I(q) - K E(q,s)\right],
\end{equation}
where $I(q)$ is the gravitational action,
\begin{equation}
I(q) = - \beta_g \sum_t A_t \delta_t + \lambda \sum_i V_i.
\end{equation}
The first sum runs over all products of triangle areas $A_t$ times corresponding
deficit angles $\delta_t$ weighted by the gravitational coupling $\beta_g$.
The second sum extends over the volumes $V_i$ of the
4-simplices of the lattice and allows one together with the cosmological constant
$\lambda$ to set an overall scale in the action.
The energy of  Ising spins $s_i  \in Z_2$,
\begin{equation}
E(q,s) = \frac{1}{2} \sum_{\langle ij \rangle} A_{ij}\frac{(s_i-s_j)^2}{q_{ij}},
\end{equation}
is defined as in two dimensions \cite{gross,holm,physA}, with the barycentric area $A_{ij}$ associated
with a link~$l_{ij}$,
\begin{equation}
A_{ij} = \sum _{\mbox{\scriptsize $t \supset l_{ij}$}}
\frac{1}{3} A_t.
\end{equation}
We chose the simple uniform measure as in the pure gravity simulations \cite{homo}:
\begin{equation}
D[q]=\prod_l dq_l {\cal F}(q_l).
\end{equation}
The function ${\cal F}$ ensures that only
Euclidean link configurations are taken into account, i.e., ${\cal F}=1$ if
the Euclidean triangle inequalities are fulfilled and ${\cal F}=0$ otherwise.
This is always guaranteed for the discrete Regge model by construction.

For every Monte Carlo simulation run we recorded the time series of the energy density
 $e=E/N_0$ and the magnetization density $m= \sum_i s_i /N_0$, with the lattice size $N_0=L^4$.
To obtain results for the various observables ${\cal O}$ at values of the spin coupling $K$
 in an interval around the simulation point $K_0$, we applied the
reweighting method~\cite{ferrenberg}. Since we recorded the time series, this amounts to computing
\begin{equation}
\langle {\cal O} \rangle |_K =\frac{\langle {\cal O}e^{-\Delta K E}\rangle|_{K_0}}
{\langle e^{-\Delta K E}\rangle|_{K_0}}~,
\end{equation}
with $\Delta K = K - K_0$.

With the help of the time series we can compute the specific heat,
\begin{equation}
C(K)=K^2 N_0 (\langle e^2 \rangle- \langle e\rangle^2)~,
\end{equation}
the (finite lattice) susceptibility,
\begin{equation}
\chi(K)=N_0(\langle m^2 \rangle -\langle |m| \rangle^2)~,
\end{equation}
and various derivatives of the magnetization,
$d \langle |m| \rangle/dK$,
$d$ln$\langle |m| \rangle
/dK$, and
$d$ln$\langle m^2 \rangle/dK$.
All these quantities exhibit in the infinite-volume limit singularities at
$K_c$ which are shifted and rounded in finite systems.
We further analyzed the Binder parameter,
\begin{equation}
U_L(K)=1-\frac{1}{3}\frac{\langle m^4 \rangle}{\langle m^2\rangle^2}~.
\end{equation}
It is well known that the $U_L(K)$ curves for different lattice sizes $L$ cross around
$(K_c,U^*)$.
This allows an almost unbiased estimate of the critical spin coupling $K_c$.

\section{Simulation results} \label{sect3}

In four dimensions, after initial discussions~\cite{gaunt79,gaunt80,velasco}
it is generally accepted that the critical properties of
the Ising model on a static lattice are given by mean-field theory
with logarithmic corrections. The finite-size scaling (FSS) formulas can
be written as~\cite{ralph,parisi}
\begin{eqnarray}
\xi &\propto& L(\log L)^{\frac{1}{4}}, \\
\chi  &\propto& L^2 (\log L)^{\frac{1}{2}} \quad = (L (\log L)^{\frac{1}{4}})^{\gamma/\nu},\\
C   &\propto& (\log L)^{\frac{1}{3}}, \\
K_c(\infty)-K_c(L) &\propto& L^{-2} (\log L)^{-\frac{1}{6}} = (L (\log L)^{\frac{1}{12}})^{-1/
\nu}, \label{e:4d_kc}
\end{eqnarray}
where the critical exponents of mean-field theory are $\alpha=0$, $\beta=1/2$,
$\gamma=1$, and $\nu=1/2$.
To get more precise ideas about these logarithmic corrections,
we first simulated the
four-dimensional Ising model on a regular lattice. After this comparative study
we turned to the four-dimensional
discrete Regge model \cite{homo} with spin fields.

\subsection{Ising spins on a regular lattice}

We studied the four-dimensional Ising model on a hypercubic lattice
 with linear size $L= 3 - 16$, 18, 20, 24, 28, 32, 36, 40, using the
single-cluster update algorithm (Wolff)~\cite{wolff}.
The simulations were performed with the knowledge of the value of the critical temperature
obtained in previous Monte Carlo simulations and high-temperature series 
analyses~\cite{staufad}:
\begin{equation}
K_c=\frac{J}{k_B T_c}=0.149\,694\pm0.000\,002.
\end{equation}

We performed $n(L) \propto N_0/\langle C(L)\rangle$ cluster updates
between measurements for lattices $L\le 24$, with the averaged
 cluster size $\langle C(L)\rangle$. After an
initial equilibration time of about $1\,000 \times n(L)$ cluster updates
we took about $50\,000$ measurements for each
of the small lattices.
For the larger lattices we measured after each cluster update,
therefore, we took about $500\,000$ measurements after an initial equilibration
of $100\,000$ cluster updates.
Analyzing the time series we found integrated autocorrelation
times for the energy and the magnetization in the range of unity for the small 
lattices $L\le 24$ and in the range of $(4-8) \times L$ for the larger lattice
sizes. The statistical errors were obtained by the standard jack-knife method
using 50 blocks.

\begin{figure*}[hp]
\psrotatefirst
\centerline{\hbox{
\psfig{figure=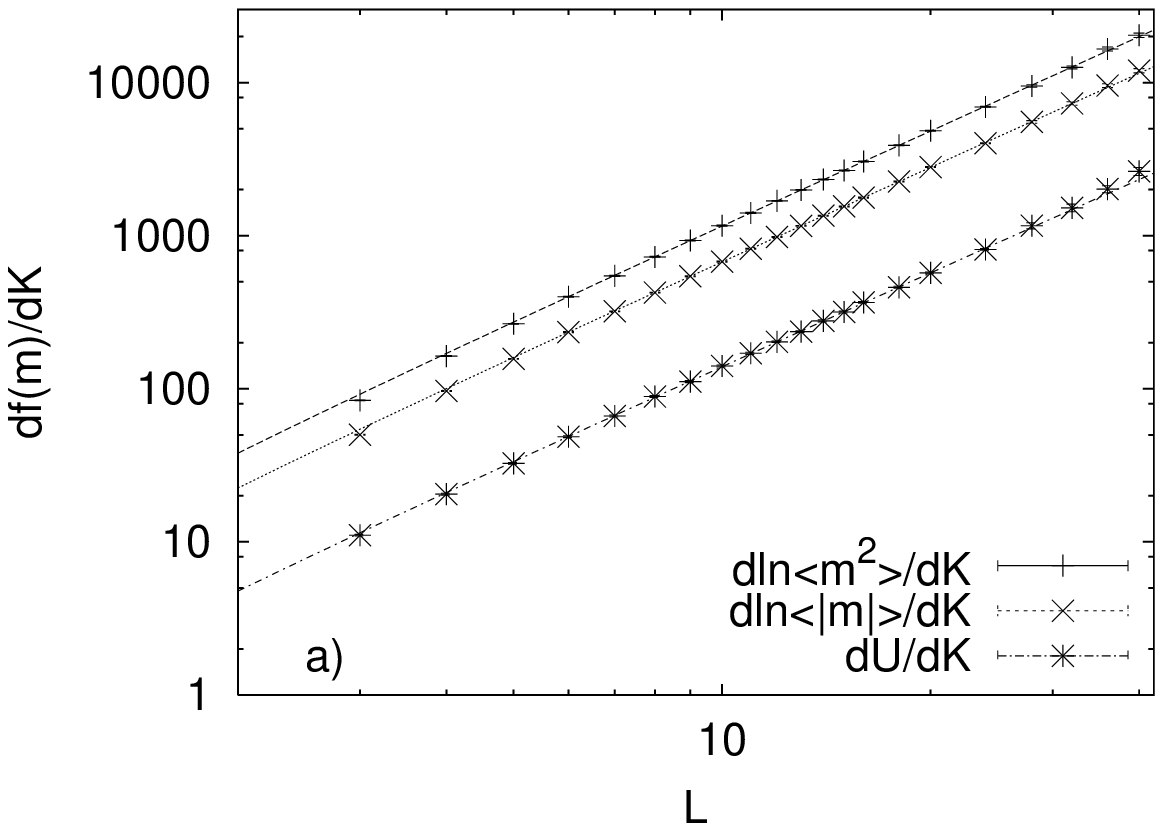,height=6.5cm,width=9.cm}
\psfig{figure=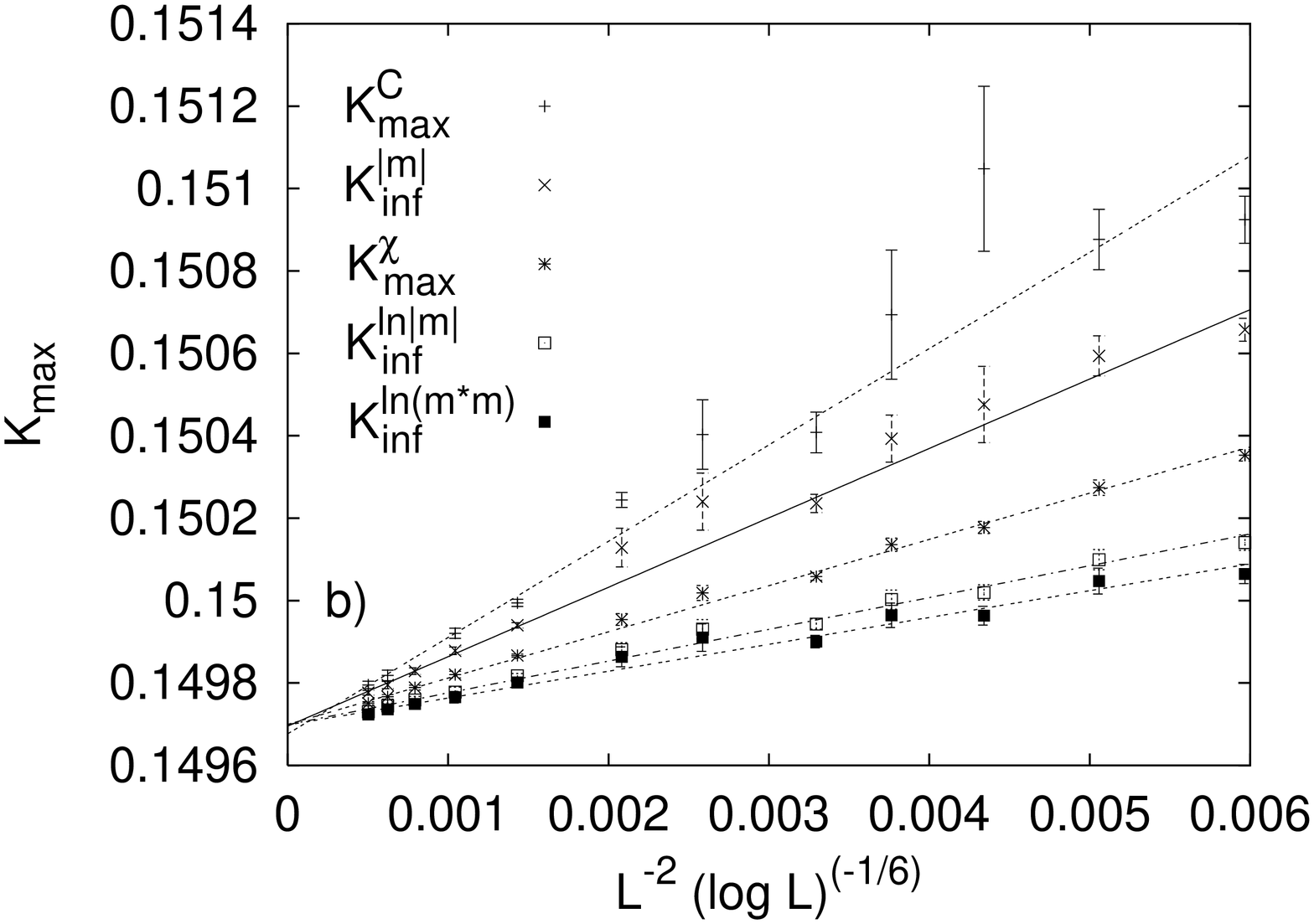,angle=-90,height=6.5cm,width=9.cm}
}}

\vspace{3mm}
\caption{(a) Least-square fits of the FSS
ansatz with logarithmic corrections at the maxima locations. Together with the fits
at $K_c$ this leads to an overall critical exponent $1/\nu=1.994(2)$.
(b) FSS extrapolations of pseudo-transition points $K_{\rm max}$ vs.
$(L (\log L)^{\frac{1}{12}})^{-1/\nu}$, assuming $\nu=0.5$. The error-weighted average of
extrapolations to infinite size yields $K_c=0.149\,697(2)$.
}
\label{kc4di}
\label{4ddU}
\end{figure*}

Applying the reweighting technique we first determined the maxima of $C$,
$\chi$,  $d\langle |m| \rangle /dK$, $d$ln$\langle |m| \rangle /dK$,
and $d$ln$\langle m^2 \rangle/dK$. The locations of the maxima provide us with
five sequences of pseudo-transition points $K_{\rm max}(L)$ for which 
the scaling variable
$x=( K_c - K_{\rm max}(L)) (L (\log L)^\frac{1}{12})^{1/\nu}$ 
should be constant.
Using this fact we then have several 
possibilities to extract the critical exponent $\nu$ from (linear)
 least-square fits of the FSS ansatz with multiplicative logarithmic
corrections considering Eq.~(\ref{e:4d_kc}), 
\begin{eqnarray}
dU_L/dK &\cong&  (L (\log L)^\frac{1}{12})^{1/ \nu} f_0(x) \quad {\rm or}\label{e:4d_duldk}\\
d{\rm ln}\langle |m|^p\rangle /dK &\cong&  (L (\log L)^\frac{1}{12})^{1/\nu} f_p(x),
\end{eqnarray}
to the data at the various $K_{\rm max}(L)$ sequences. 
For comparison we also performed fits of a naive power-law FSS ansatz
\begin{eqnarray}
dU_L/dK &\cong&  L^{1/ \nu} f_0(x) \quad {\rm or}\\
d{\rm ln}\langle |m|^p\rangle /dK &\cong&  L^{1/\nu} f_p(x).\label{e:4d_dlnmdk}
\end{eqnarray}
The exponents $1/\nu$ resulting from fits
using the data for the $N$ largest lattice sizes are collected in Tables~I and II. 
$Q$ denotes the standard goodness-of-fit parameter.
For all exponent estimates the FSS ansatz with the logarithmic corrections
leads to the  weighted average $1/\nu = 1.993(3)$, which is in perfect agreement with the
mean-field value $1/\nu=2$, see \hbox{Fig.~\ref{4ddU}~(a)}.
With the naive power-law ansatz one also gets an estimate for $1/\nu$ in the vicinity of
the mean-field value, but this is clearly separated from the mean-field result,
verifying the significance of the multiplicative logarithmic correction, cf. Table~II.

Assuming therefore $\nu=0.5$ we can obtain estimates for $K_c$ from linear least-square fits to 
the scaling behavior of the various $K_{\rm max}$ sequences, as shown in \hbox{Fig.~\ref{kc4di}~(b)}. 
Using the fits with $L \ge 12$, the combined estimate from 
the five sequences leads to $K_c=0.149\,697(2)$, which is in agreement with 
the results using 
Monte Carlo computer simulations \cite{staufad} and
series expansions \cite{gaunt,adstauf}. 

Knowing the critical coupling we may reconfirm our estimates of $1/\nu$ by
evaluating the above quantities at $K_c$. As can be seen in Tables~I and II,
the statistical errors of the 
FSS fits at $K_c$ are similar to those using the $K_{\rm max}$ 
sequences.  However, the uncertainty in the estimate of $K_c$ has also to be taken into
account. This error is computed by repeating the
fits at $K_c \pm \Delta K_c$ and indicated in Tables~I and II by the numbers in 
square brackets. In the computation of the weighted average we assume the two
types of errors to be independent.
As a result of this combined analysis we obtain strong evidence that the exponent 
$\nu$ agrees with the mean-field value of $\nu=0.5$.

\begin{figure*}[hp]
\centerline{\hbox{
\psfig{figure=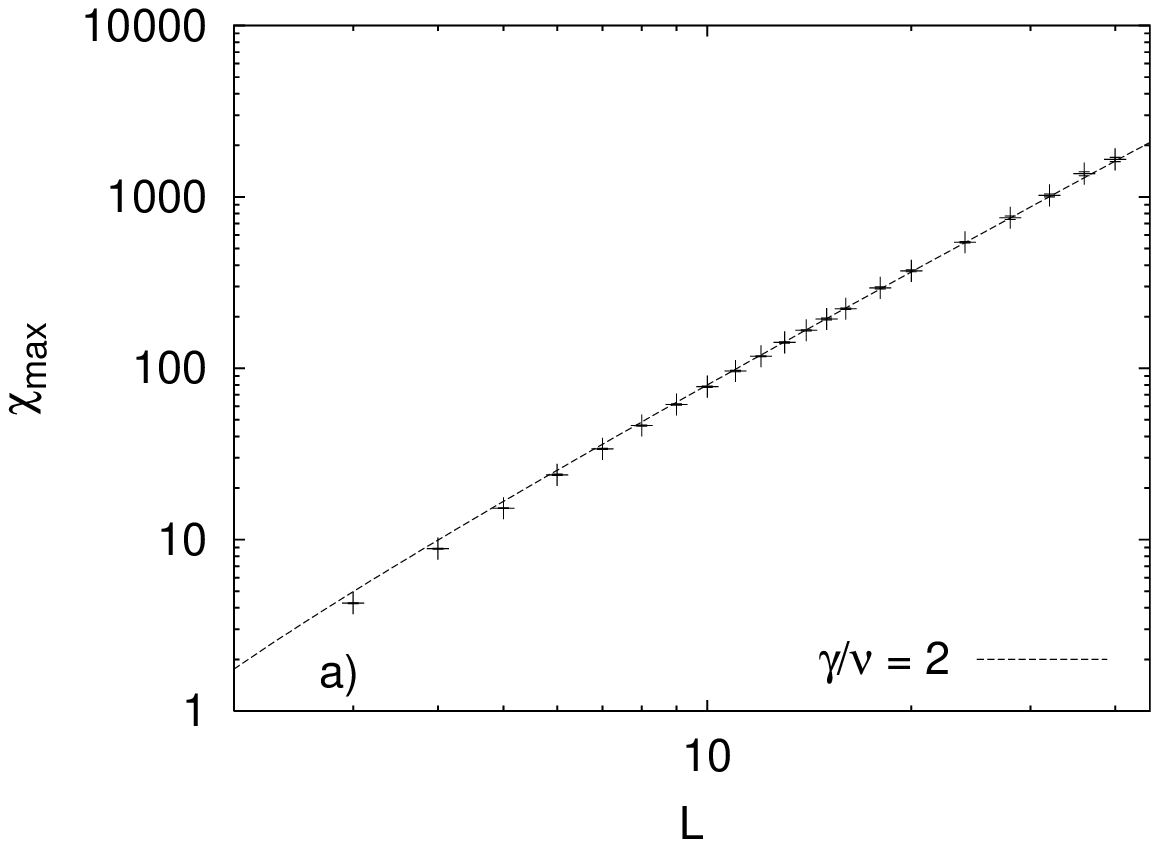,height=6.5cm,width=9.cm}
\psfig{figure=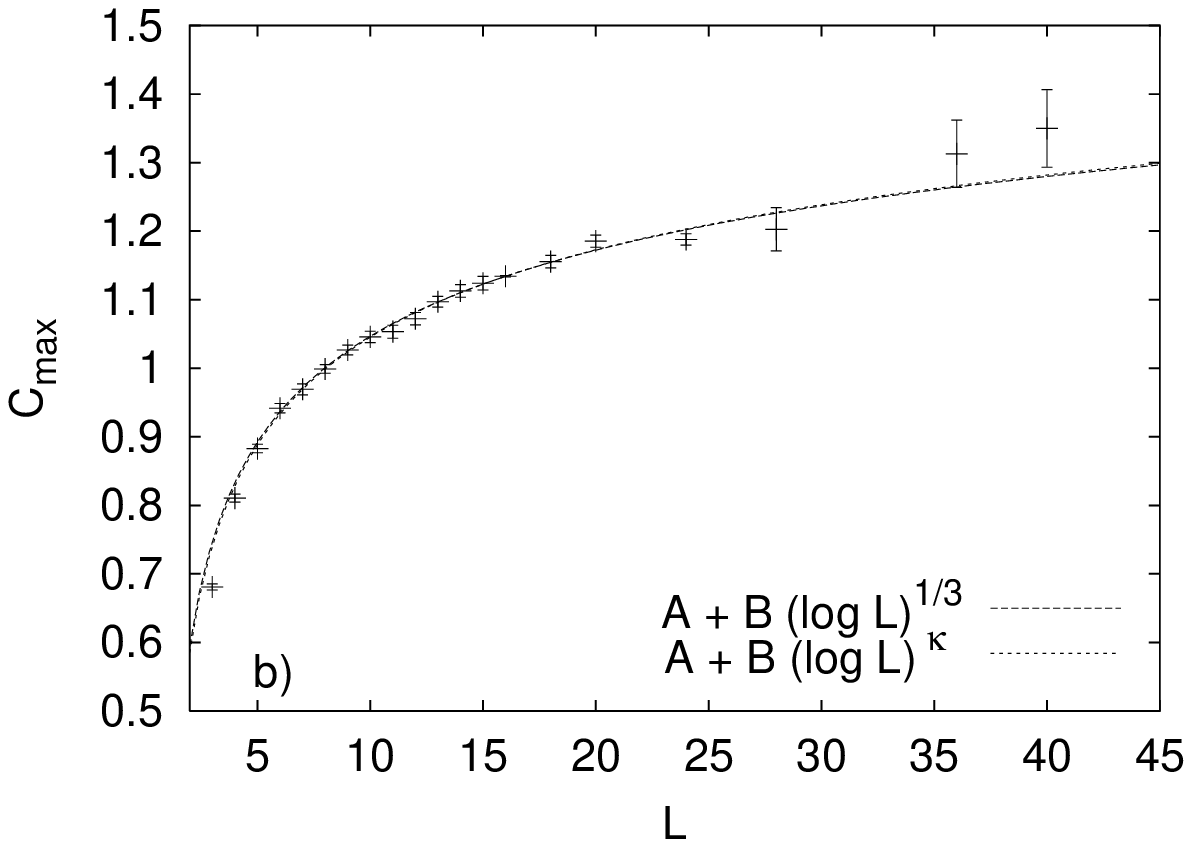,height=6.5cm,width=9.cm}
}}

\vspace{3mm}
\caption{
(a) FSS of the susceptibility maxima $\chi_{\rm max}$.
The exponent entering the curve is set to the mean-field value
$\gamma / \nu = 2$ for regular static lattices.
(b) FSS of the specific-heat maxima $C_{\rm max}$.
The logarithmic fit \hbox{$C_{\rm max} = A + B (\log L)^{\kappa} $}
and a constrained logarithmic fit assuming the mean-field prediction
\hbox{$\kappa = 1/3$} are almost indistinguishable on the
scale of the figure.
}
\label{fig_c4di}
\label{fig_sus4di}
\end{figure*}

To extract the critical exponent ratio $\gamma/\nu$ 
we use the scaling 
\begin{equation}
\chi_{\rm max} \cong (L (\log L)^{\frac{1}{4}})^{\gamma/\nu}
\label{4d:sus_fit}
\end{equation}
as well as the scaling of $\chi$ 
at $K_c$, yielding in the range $L=10 - 40$ estimates of 
$\gamma/\nu=2.037(9)$ with $Q=0.95$ and 
$\gamma/\nu=2.008(5)[5]$ ($Q=0.46$), respectively. 
These estimates for $\gamma/\nu$ are consistent with the mean-field value of 
$\gamma/\nu=2$. In \hbox{Fig.~\ref{fig_sus4di}~(a)} this is demonstrated graphically
by comparing the scaling of $\chi_{\rm max}$ with a constrained one-parameter fit of 
the form $\chi_{\rm max} = c (L (\log L)^{\frac{1}{4}})^{2}$ with 
$c=0.526(2)$ ($Q=0.38$, $L \ge 6$).

\begin{figure*}[hp]
\centerline{\hbox{
\psfig{figure=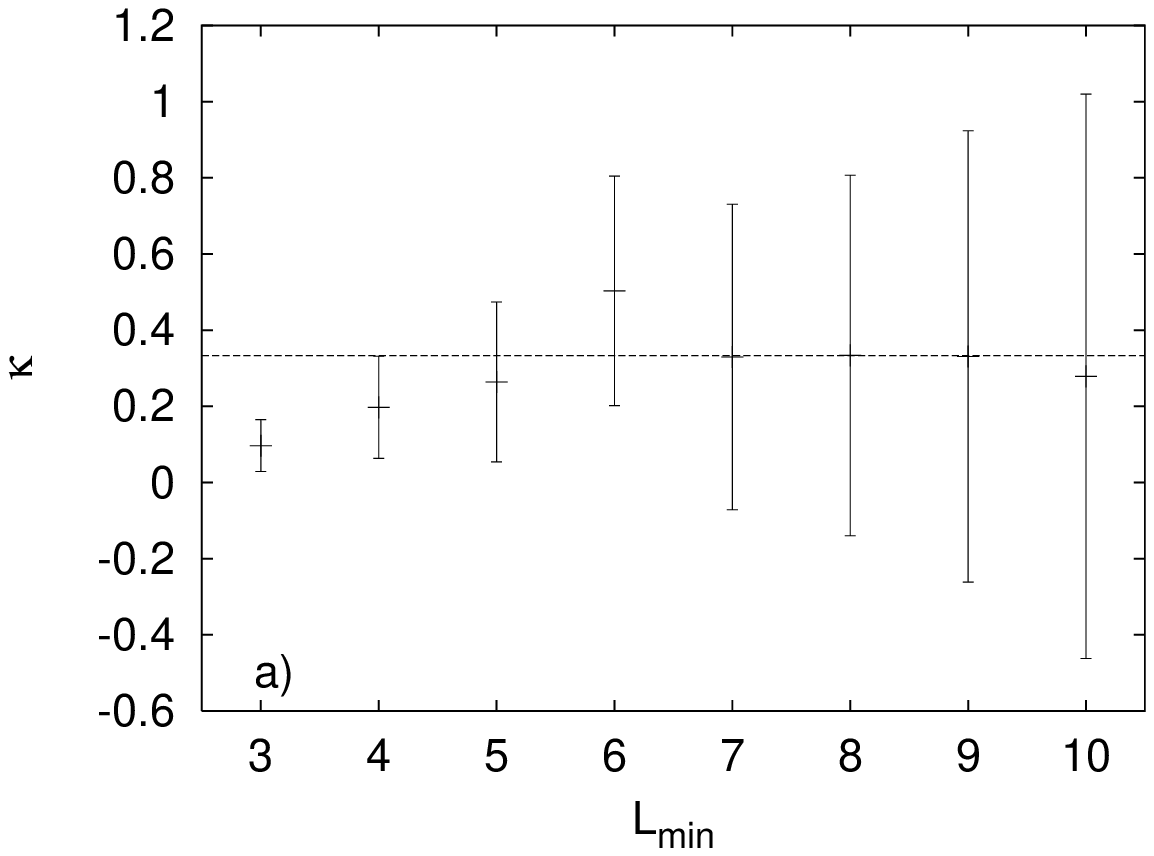,height=6.5cm,width=9.cm}
\psfig{figure=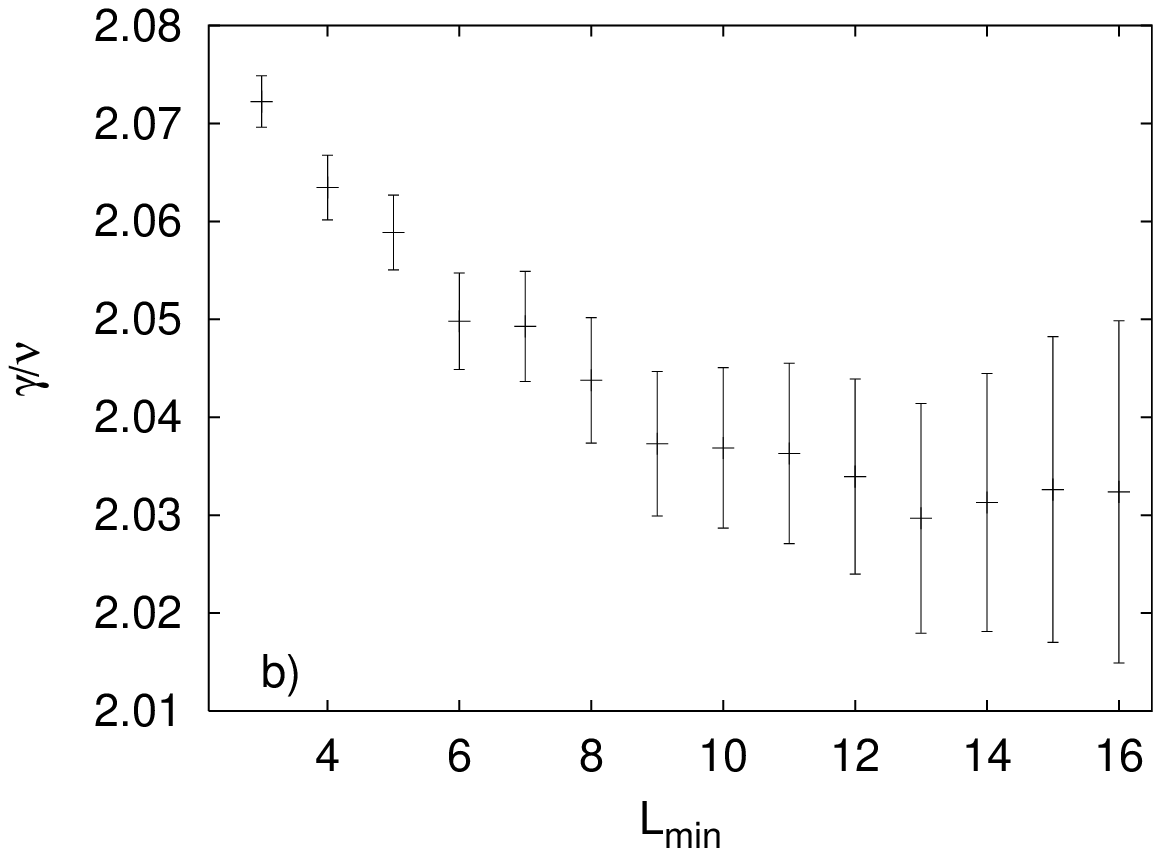,height=6.5cm,width=9.cm}
}}

\vspace{3mm}
\caption{
(a) Instability of the logarithmic fit
$C_{\rm max} = A + B (\log L)^{\kappa}$.
The exponent $\kappa$ as a function of the lower bound of the fit range is plotted.
(b) $\gamma/\nu$ as a function of the lower bound of the fit range is plotted for
the fit $\chi_{\rm max} \propto  (L(\log L)^\frac{1}{4})^{\gamma/\nu}$.
}
\label{fig_sus_E4di}
\label{fig_c_E4di}
\end{figure*}

Concerning the specific heat we expect in the case of the mean-field exponent 
$\alpha=0$ a logarithmic divergence of the form
\begin{equation}
C(x,L)=A(x)+B(x)(\log L)^{\frac{1}{3}}.
\label{eq:C_log_4d}
\end{equation}
Indeed, the data at the different fixed values of $x$ can all be fitted nicely with this ansatz. 
In particular, for the fit of $C_{\rm max}$ with 17 points ($L \ge 6$) 
we obtain $A=-0.324(32)$, $B=1.038(23)$, with a total $\chi^2=11.7 (Q=0.70)$.
We also tried an unbiased three-parameter fit using the ansatz
\begin{equation}
C(x,L)=A'(x)+B'(x)(\log L)^{\kappa(x)},
\label{eq:C_3para_4d}
\end{equation}
which in the case of the fit of $C_{\rm max}$ and 16 data points gives
$A' \approx  -0.36$,      %
$B' \approx  1.75$,      %
and $\kappa=0.33(40)$,      %
with a slightly improved total $\chi^2=10.8 (Q=0.62)$. 
In \hbox{Fig.~\ref{fig_c4di}~(b)} we compare these two linear least-square fits.
It should be noted, however, that the three-parameter fit is highly unstable 
and exhibits strong correlations between the three parameters. 
To illustrate this instability we plot in \hbox{Fig.~\ref{fig_c_E4di}~(a)} the exponent $\kappa$
as a function of the
smallest lattice size $L_{\rm{min}}$, being the lower bound of the fit range
$[L_{\rm{min}},40]$. For comparison we show in \hbox{Fig.~\ref{fig_sus_E4di}~(b)}
the behavior of  $\gamma/\nu$ results of the fit corresponding to Eq.~(\ref{4d:sus_fit}).

\begin{center}
\begin{tabular}{|l|c|l|c|}\hline
\makebox[4cm][c]{fit type} & \makebox[1cm][c]{$N$} &\makebox[2.cm][c]{ $1/\nu$}
 &\makebox[2cm][c]{$Q$}\\ \hline
$dU/dK$ at $K^C_{\rm max}$ & 22 &~1.990(5)& 0.68\\
$d$ln$\langle |m|\rangle /dK$ at $K^{\ln\langle |m|\rangle}_{\rm inf}$ & 18 &~1.989(4)&0.94 \\
$d$ln$\langle m^2\rangle/dK$ at $K^{\ln\langle m^2\rangle}_{\rm inf}$ & 18 &~1.998(5)&0.93 \\
weighted average && ~1.993(3)&\\ \hline
$dU/dK$ at $K_c$ & 18 &~1.991(5)[1]&0.73 \\
$d$ln$\langle |m|\rangle /dK$ at $K_c$ & 18 &~1.992(5)[2]& 0.94\\
$d$ln$\langle m^2\rangle/dK$ at $K_c$ & 18 &~2.000(5)[2]& 0.94\\
weighted average && ~1.995(3)&\\ \hline
\hline
overall average &&  ~1.994(2)&\\ \hline
\end{tabular}
\end{center}

\vspace{3mm}
Table I. Fit results for  $1/\nu$ with a power-law ansatz with 
logarithmic corrections, using the data for the $N$ largest lattices.
The average is computed by weighting each entry with its inverse squared error.
For the fits at our best estimate of $K_c=0.149\,697(2)$ the uncertainty
due to the error in $K_c$ is indicated by the numbers in square brackets.

\begin{center}
\begin{tabular}{|l|c|l|c|}\hline
\makebox[4cm][c]{fit type} & \makebox[1cm][c]{$N$} &\makebox[2.cm][c]{ $1/\nu$}
&\makebox[2cm][c]{$Q$}\\ \hline
$dU/dK$ at $K^C_{\rm max}$ & 18 &~2.041(9)& 0.81 \\
$d$ln$\langle |m|\rangle /dK$ at $K^{\ln\langle |m|\rangle}_{\rm inf}$ & 18 &~2.056(5)&0.75 \\
$d$ln$\langle m^2\rangle/dK$ at $K^{\ln\langle m^2\rangle}_{\rm inf}$ & 18 &~2.066(5)&0.63 \\
weighted average && ~2.059(3)&\\ \hline
$dU/dK$ at $K_c$ & 18 &~2.043(9)[2]&0.81 \\
$d$ln$\langle |m|\rangle /dK$ at $K_c$ & 18 &~2.059(5)[2]& 0.81\\
$d$ln$\langle m^2\rangle/dK$ at $K_c$ & 18 &~2.067(5)[2] &0.70\\
weighted average && ~2.061(4)&\\ \hline
\hline
overall average &&  ~2.060(2)&\\ \hline
\end{tabular}
\end{center}

\vspace{3mm}
Table II. Fit results for  $1/\nu$ with a pure power-law
ansatz using $K_c=0.149\,697(2)$.
The averages and statistical errors are computed as in Table~I.


\subsection{Ising spins on a discrete Regge model}

We simulated the gravitational degrees of freedom (the squared link lengths)
of the partition function (\ref{z}) using the heat-bath algorithm.
For the Ising spins we employed again the single-cluster algorithm.
Between measurements we performed $n=10$ Monte Carlo steps consisting of one
lattice sweep to update the squared link lengths~$q_{ij}$
followed by two single-cluster flips to
update the spins~$s_i$.

\begin{figure*}[hp]
\psrotatefirst
\centerline{\hbox{
\psfig{figure=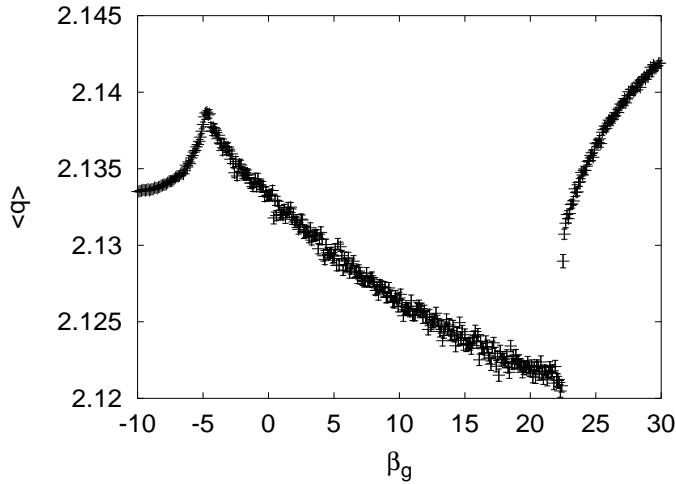,angle=-90,height=6.5cm,width=9.5cm}
}}

\vspace{3mm}
\caption{Expectation values of the average squared link lengths as a function of the
gravitational coupling for the pure discrete Regge model on a $4^4$ lattice.
}
\label{4dz2_q}
\end{figure*}

\begin{figure*}[hp]
\psrotatefirst
\centerline{\hbox{
\psfig{figure=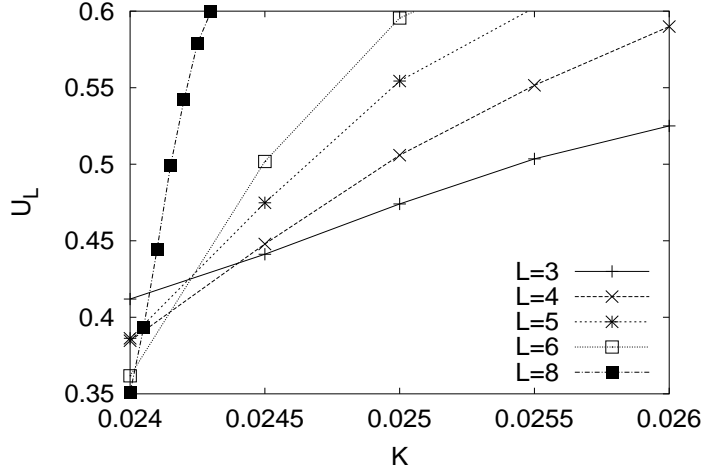,height=6.5cm,width=9.5cm}
}}

\vspace{3mm}
\caption{
Binder cumulant curves from the short runs at $\beta_g=-4.665$ leading to a critical
spin coupling $K_c \approx K_0 = 0.024$.}
\label{4dU_short}
\end{figure*}

The simulations were done
for $\epsilon=0.0875$, cosmological constant $\lambda = 0$ and two
different gravitational couplings, $\beta_g=-4.665$ and $\beta_g=22.3$. 
These two $\beta_g$-values correspond to the two 
phase transitions of the pure discrete Regge model~\cite{homo}, as shown in Fig.~\ref{4dz2_q}.
The transition at positive gravitational coupling of the standard Regge calculus
was described in great detail in Ref.~\cite{hamber00} and shown to be of
first order whereas the transition at negative coupling turned out to be
of second order for the discrete Regge model~\cite{homo}. Together with
an eventual second-order transition of the Ising part, the latter one is
a candidate for a possible continuum limit.
The lattice topology is given by triangulated tori of size $N_0=L^4$ with
$L = 3$ up to 10.
 From short test runs we estimated the location of the phase
transition of the spin model and set the spin coupling $K_0 = 0.024 \approx K_c$ in the long runs for
both values of $\beta_g$, see  Fig.~\ref{4dU_short}.

After an initial equilibration time we took about 
$100\,000$ measurements
for each lattice size. Analyzing the time series we found integrated
 autocorrelation
times for the energy and the magnetization in the range of unity for all 
considered lattice sizes. As in the simulations of the regular lattices
the statistical errors were obtained by the standard
jack-knife method using 50 blocks.

\begin{figure*}[hp]
\psrotatefirst  
\centerline{\hbox{
\psfig{figure=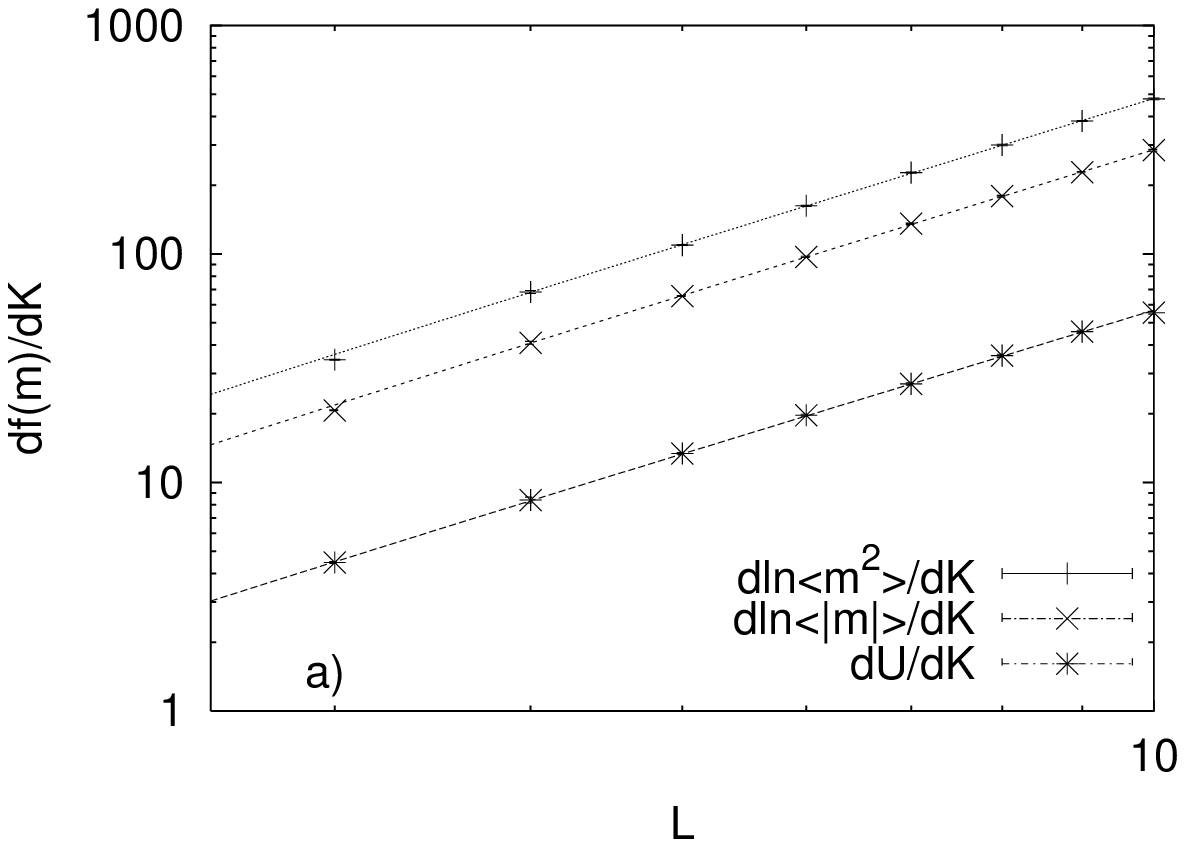,height=6.5cm,width=9.5cm}
\psfig{figure=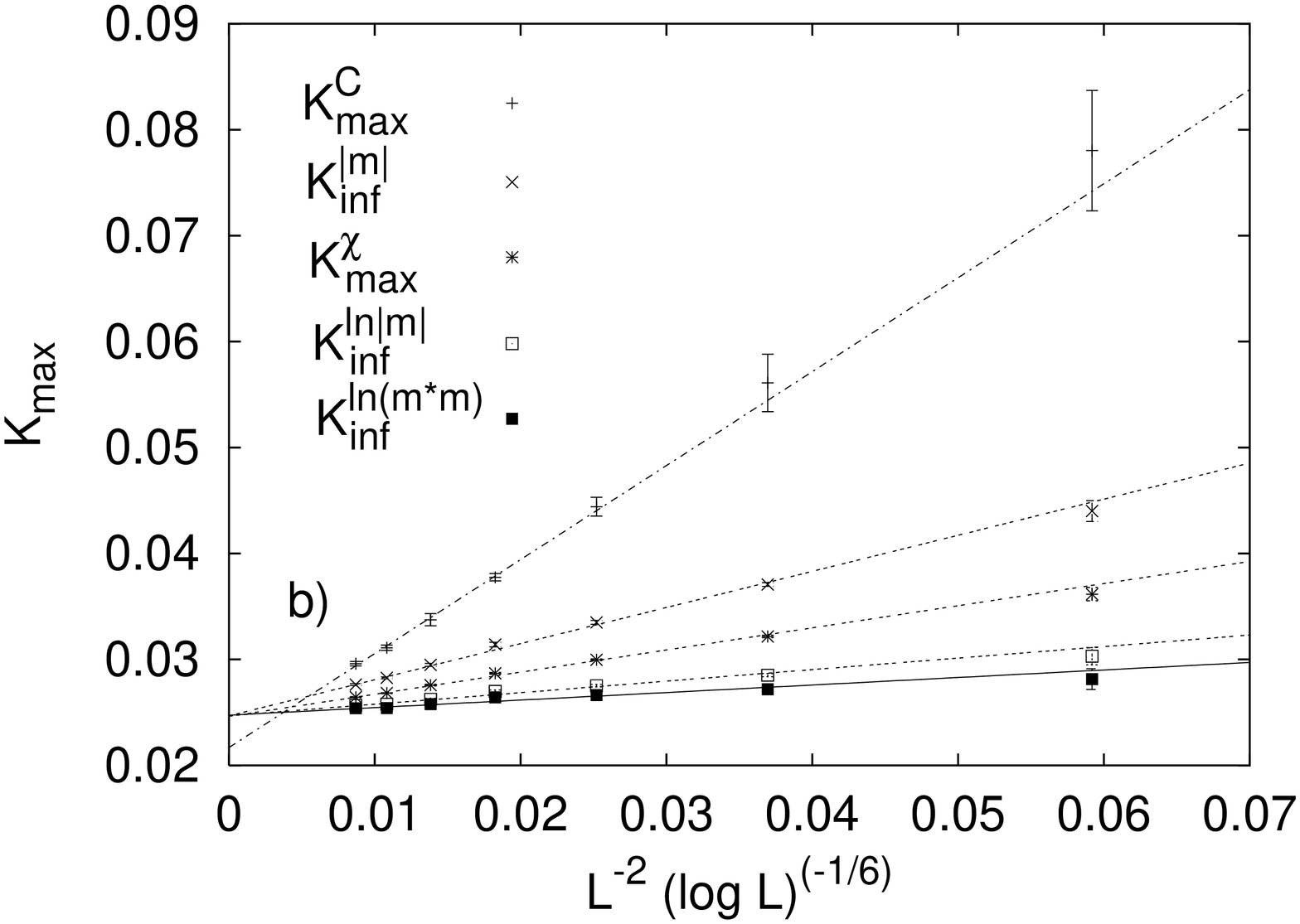,angle=-90,height=6.5cm,width=9.5cm}
}}

\vspace{3mm}
\caption{(a) Least-square fits of the FSS
ansatz with logarithmic corrections for $\beta_g=-4.665$
 lead to an overall critical exponent $1/\nu=2.025(4)$.
(b) FSS extrapolations of pseudo-transition points $K_{\rm max}$ vs.
$(L (\log L)^{\frac{1}{12}})^{-1/\nu}$ for $\beta_g=-4.665$, assuming $\nu=0.5$.
 The error-weighted average of
extrapolations to infinite size yields $K_c=0.02464(4)$.
}
\label{kc4diz2}
\label{4ddUz2}
\end{figure*}

Completely analogously to the Ising system on a regular lattice we applied
reweighting to locate the maxima and used the FSS 
formulas~(\ref{e:4d_duldk})--(\ref{e:4d_dlnmdk}).
The exponents $1/\nu$ resulting from fits
using the data for the $N$ largest lattice sizes are collected in Tables~III and IV
for $\beta_g=-4.665$, and in Tables~V and VI for $\beta_g=22.3$, respectively. 
For the simulations at $\beta_g=-4.665$ all exponent estimates with the logarithmic corrections
and consequently also their weighted average $1/\nu = 2.025(6)$ 
are in agreement with the mean-field value $1/\nu=2$, see \hbox{Fig.~\ref{4ddUz2}~(a)}.
For $\beta_g=22.3$ the scatter in the estimates is similar and the weighted 
average
$1/\nu = 2.028(6)$ is again compatible with $1/\nu = 2$. 
With the naive power-law ansatz one also gets an estimate for $1/\nu$ in the vicinity of
the mean-field value, but this is clearly separated from the mean-field result, cf. Tables~IV
and VI.

Assuming therefore $\nu=0.5$ we can obtain estimates for $K_c$ from linear least-square fits to 
the scaling behavior of the various $K_{\rm max}$ sequences, as shown in hbox{Fig.~\ref{kc4diz2}~(b)} for
$\beta_g=-4.665$. 
Using the fits with $L \ge 4$, the combined estimate from 
the five sequences leads to $K_c=0.02464(4)$ for $\beta_g=-4.665$, and for $\beta_g=22.3$
 we find $K_c=0.02339(4)$, again with $L \ge 4$.

With the knowledge of the critical couplings we may reconfirm our estimates of $1/\nu$ by
evaluating the above quantities at $K_c$. As can be inspected in Tables~III and V,
we obtain from this combined analysis strong evidence that the exponent 
$\nu$ agrees with the mean-field value of $\nu=0.5$.

\begin{figure*}[hp]
\centerline{\hbox{
\psfig{figure=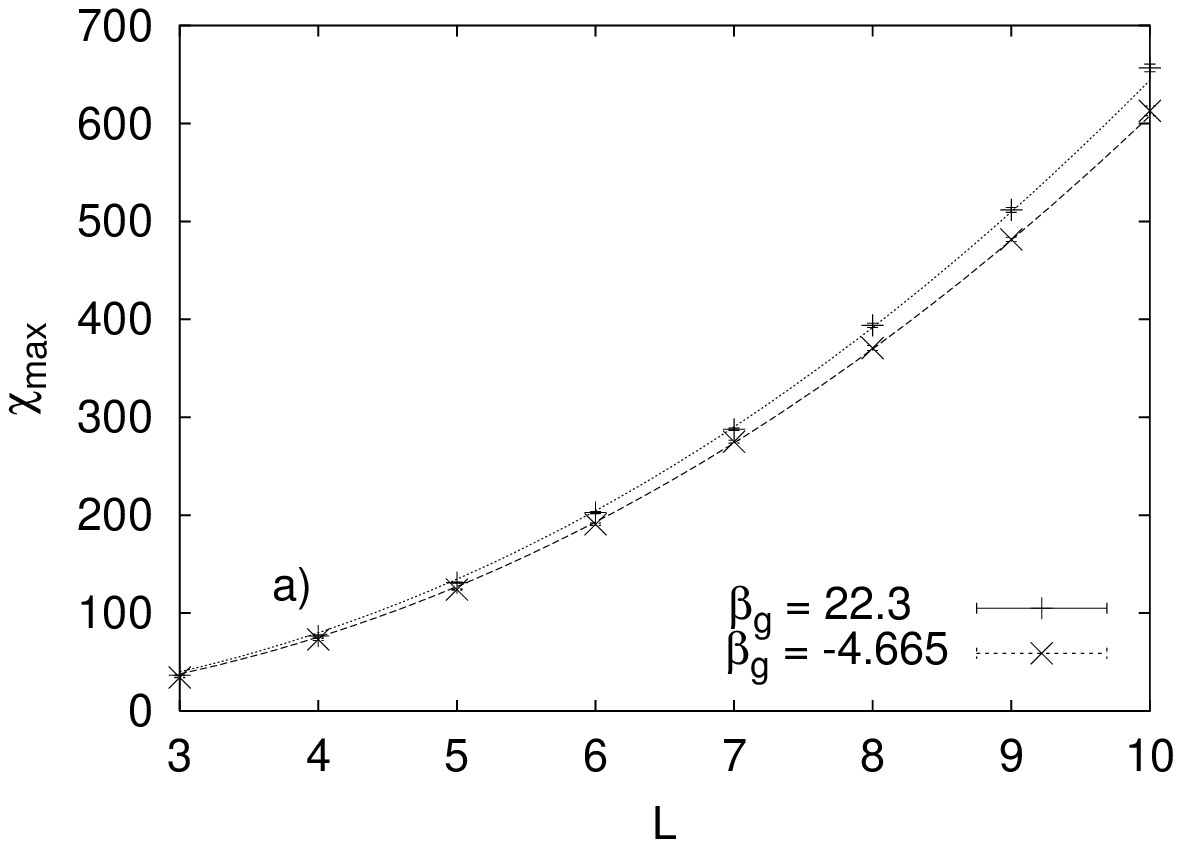,height=6.5cm,width=9.5cm}
\psfig{figure=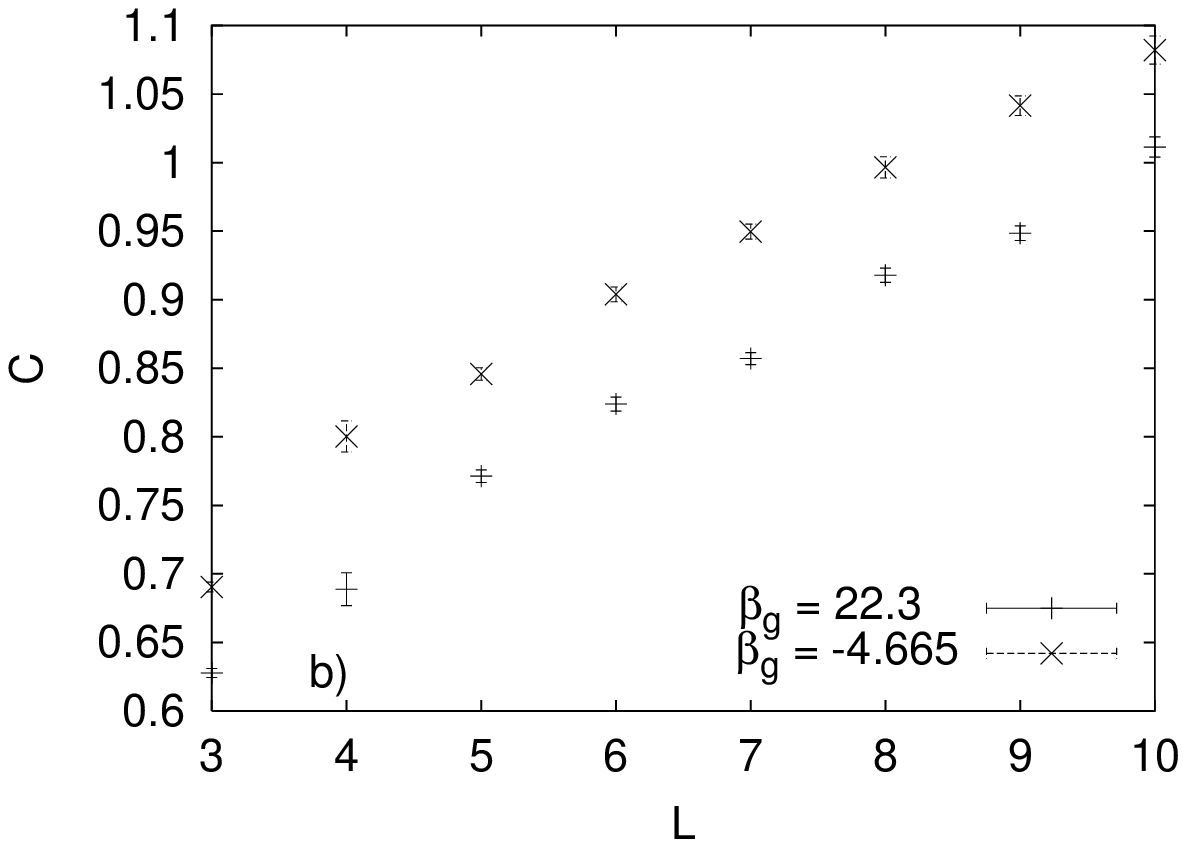,height=6.5cm,width=9.5cm}
}}

\vspace{3mm}
\caption{
(a) FSS of the susceptibility maxima $\chi_{\rm max}$.
The exponent entering the curve is set to the mean-field value $\gamma / \nu = 2$ for
regular static lattices.
(b) Specific heat at the critical spin coupling as a 
function of the lattice size $L$.
}
\label{4dcz2}
\label{fig_sus4diz2}
\end{figure*}

To extract the critical exponent ratio $\gamma/\nu$ 
we use the FSS formula~(\ref{4d:sus_fit}) for $\chi_{\rm max}$
as well as the scaling of $\chi$ 
at $K_c$, yielding for $\beta_g=-4.665$ in the range $L=4 - 10$
 estimates of $\gamma/\nu=2.039(9)$ with $Q=0.42$ and
$\gamma/\nu=2.036(7)[4]$ ($Q=0.85$), respectively. 
The corresponding values for $\beta_g=22.3$,
using the same fit range, are $\gamma/\nu=2.052(8)$ ($Q=0.57$) and 
$\gamma/\nu= 2.052(6)[4]$ ($Q=0.01$). 
These estimates for $\gamma/\nu$ are compatible with the mean-field value of 
$\gamma/\nu=2$.
In \hbox{Fig.~\ref{fig_sus4diz2}~(a)} this is demonstrated
by comparing the scaling of $\chi_{\rm max}$ with a constrained one-parameter fit of 
the form $\chi_{\rm max} = c (L (\log L)^{\frac{1}{4}})^{2}$ with 
$c=4.006(10)$ ($Q=0.17$, $L \ge 6$) for $\beta_g=-4.665$ and $c= 4.244(10)$ ($Q=0.001$, $L \ge 6$) for
$\beta_g=22.3$, respectively.

The data for the specific heat $C$ at the critical spin coupling $K_c$ 
are presented in \hbox{Fig.~\ref{4dcz2}~(b)}. The fact that $C$ increases very slowly
with the size of the lattice means that one will need data from bigger lattices and more
statistical accuracy to get an estimate or bound for the critical exponent $\alpha$
from a direct measurement of $C$.
Especially, if we assume a logarithmic divergence of $C$ as in the four-dimensional
Ising model on regular lattices, we need lattices of comparable size,
cf. \hbox{Fig.~\ref{fig_c4di}~(b)}. 

\vspace{5mm}


\begin{center}
\begin{tabular}{|l|c|l|c|}\hline
\makebox[4cm][c]{fit type} & \makebox[1cm][c]{$N$} &\makebox[2.cm][c]{ $1/\nu$}
&\makebox[2cm][c]{$Q$}\\ \hline
$dU/dK$ at $K^C_{\rm max}$ & 8 &~2.003(10)& 0.47\\
$d$ln$\langle |m|\rangle /dK$ at $K^{\ln\langle |m|\rangle}_{\rm inf}$ & 7 &~2.032(10)&0.59 \\
$d$ln$\langle m^2\rangle/dK$ at $K^{\ln\langle m^2\rangle}_{\rm inf}$ & 7 &~2.038(10)&0.55 \\
weighted average && ~2.025(6)&\\ \hline
$dU/dK$ at $K_c$ & 7 &~1.981(17)[13]&0.70 \\
$d$ln$\langle |m|\rangle /dK$ at $K_c$ & 7 &~2.027(9)[2]&0.95\\
$d$ln$\langle m^2\rangle/dK$ at $K_c$ & 7 &~2.034(9)[2]& 0.85\\
weighted average && ~2.025(6)&\\ \hline
\hline
overall average &&  ~2.025(4)&\\ \hline
\end{tabular}
\end{center}

\vspace{3mm}
Table III. Fit results for  $1/\nu$ with a power-law ansatz with 
logarithmic corrections for $\beta_g=-4.665$, using the data for the $N$
largest lattices.
The average is computed by weighting each entry with its inverse squared error.
For the fits at our best estimate of $K_c=0.02464(4)$ the uncertainty
due to the error in $K_c$ is indicated by the numbers in square brackets.

\vspace{3mm}
\begin{center}
\begin{tabular}{|l|c|l|c|}\hline
\makebox[4cm][c]{fit type} & \makebox[1cm][c]{$N$} &\makebox[2.cm][c]{ $1/\nu$}
&\makebox[2cm][c]{$Q$}\\ \hline
$dU/dK$ at $K^C_{\rm max}$ & 7 &~2.068(18)& 0.60 \\
$d$ln$\langle |m|\rangle /dK$ at $K^{\ln\langle |m|\rangle}_{\rm inf}$ & 7 &~2.122(10)&0.37 \\
$d$ln$\langle m^2\rangle/dK$ at $K^{\ln\langle m^2\rangle}_{\rm inf}$ & 7 &~2.128(10)&0.35 \\
weighted average && ~2.118(7)&\\ \hline
$dU/dK$ at $K_c$ & 7 &~2.068(18)[12]&0.59 \\
$d$ln$\langle |m|\rangle /dK$ at $K_c$ & 7 &~2.116(9)[2]& 0.83\\
$d$ln$\langle m^2\rangle/dK$ at $K_c$ & 7 &~2.124(9)[2] &0.64\\
weighted average && ~2.116(7)&\\ \hline
\hline
overall average &&  ~2.117(5)&\\ \hline
\end{tabular}
\end{center}

\vspace{3mm}
Table IV. Fit results for  $1/\nu$ with a pure power-law ansatz for $\beta_g=-4.665$.
The averages and statistical errors are computed as in Table~III.

\vspace{3mm}
\begin{center}
\begin{tabular}{|l|c|l|c|}\hline
\makebox[4cm][c]{fit type} & \makebox[1cm][c]{$N$} &\makebox[2.cm][c]{ $1/\nu$}
&\makebox[2cm][c]{$Q$}\\ \hline
$dU/dK$ at $K^C_{\rm max}$ & 8 &~1.981(10)& 0.64\\
$d$ln$\langle |m|\rangle /dK$ at $K^{\ln\langle |m|\rangle}_{\rm inf}$ & 7 &~2.043(9)&0.61 \\
$d$ln$\langle m^2\rangle/dK$ at $K^{\ln\langle m^2\rangle}_{\rm inf}$ & 7 &~2.049(9)&0.67 \\
weighted average && ~2.028(6)&\\ \hline
$dU/dK$ at $K_c$ & 8 &~1.993(10)[1]&0.76 \\
$d$ln$\langle |m|\rangle /dK$ at $K_c$ & 7 &~2.039(9)[2]&0.32\\
$d$ln$\langle m^2\rangle/dK$ at $K_c$ & 7 &~2.045(9)[2]& 0.49\\
weighted average && ~2.027(6)&\\ \hline
\hline
overall average &&  ~2.028(4)&\\ \hline
\end{tabular}
\end{center}

\vspace{3mm}
Table V. Fit results for  $1/\nu$ with a power-law ansatz with
logarithmic corrections for $\beta_g=22.3$.
The average is computed by weighting each entry with its inverse squared error.
For the fits at our best estimate of $K_c=0.02339(4)$ the uncertainty
due to the error in $K_c$ is indicated by the numbers in square brackets.

\vspace{3mm}
\begin{center}
\begin{tabular}{|l|c|l|c|}\hline
\makebox[4cm][c]{fit type} & \makebox[1cm][c]{$N$} &\makebox[2.cm][c]{ $1/\nu$}
&\makebox[2cm][c]{$Q$}\\ \hline
$dU/dK$ at $K^C_{\rm max}$ & 7 &~2.086(11)& 0.72 \\
$d$ln$\langle |m|\rangle /dK$ at $K^{\ln\langle |m|\rangle}_{\rm inf}$ & 7 &~2.134(10)&0.54 \\
$d$ln$\langle m^2\rangle/dK$ at $K^{\ln\langle m^2\rangle}_{\rm inf}$ & 7 &~2.139(9)&0.59 \\
weighted average && ~2.122(6)&\\ \hline
$dU/dK$ at $K_c$ & 8 &~2.098(10)[1]&0.57 \\
$d$ln$\langle |m|\rangle /dK$ at $K_c$ & 7 &~2.130(9)[2]& 0.35\\
$d$ln$\langle m^2\rangle/dK$ at $K_c$ & 7 &~2.135(9)[2] &0.48\\
weighted average && ~2.122(6)&\\ \hline
\hline
overall average &&  ~2.122(4)&\\ \hline
\end{tabular}
\end{center}

\vspace{3mm}
Table VI. Fit results for  $1/\nu$ with a pure power-law ansatz for $\beta_g=22.3$.
The averages and statistical errors are computed as in Table~V. 


\newpage
\section{Conclusions} \label{sect4}

We have performed a study of the Ising model coupled to fluctuating manifolds via
Regge calculus. Analyzing the discrete Regge model with two permissible edge lengths it turns out
that the Ising transition shows the predicted
 logarithmic corrections to the mean-field theory.
The critical exponents of the phase transition of the Ising spins on a static lattice 
as well as on a discrete Regge skeleton \cite{berlin} are consistent with the exponents of
the mean-field theory, $\alpha=0$, $\beta=\frac{1}{2}$, $\gamma=1$,
and $\nu=\frac{1}{2}$.
In summary, our consistent analysis with uniform computer codes yields
that the phase transition of a spin system coupled to
a discrete Regge skeleton exhibits the same critical exponents and the same
logarithmic corrections \cite{parisi} as on a static lattice.

\section*{Acknowledgments}
E.B. was supported by Fonds zur F\"orderung der wissenschaftlichen Forschung
under project P14435-TPH and thanks the Graduiertenkolleg
``Quantenfeldtheorie: Mathemati\-sche Struktur
und Anwendungen in der Elementarteilchen- und Festk\"orperphysik'' for
hospitality during his extended stay in Leipzig. W.J. acknowledges partial
support by the EC IHP Network grant HPRN-CT-1999-00161: ``EUROGRID''.



\parskip1.2ex
\begin{thebibliography}{99}
%
\bibitem{hamber84}
H.W. Hamber, in Proceedings of the 1984 Les Houches Summer School,
edited by K. Osterwalder and R. Stora, Session XLIII (North-Holland, Amsterdam, 1986);
H.W. Hamber and R.M. Williams, Phys. Lett. B {\bf 157}, 368 (1985).

%
\bibitem{berg}
B.A. Berg, in {\it Particle Physics and Astrophysics\/},
Proceedings of the XXVII Int. Universit\"atswochen f\"ur Kernphysik,
edited by H. Mitter and F. Widder (Springer, Berlin, 1989); 
Phys. Rev. Lett. {\bf 55}, 904 (1985); Phys. Lett. B {\bf 176}, 39 (1986).
%
\bibitem{beirl94}
W. Beirl, E. Gerstenmayer, H. Markum, and J. Riedler, Phys. Rev. D {\bf 49}, 
5231 (1994).
%
\bibitem{beirlz2}
W. Beirl, H. Markum, and J. Riedler, Int. J. Mod. Phys. C {\bf 5}, 359 (1994) 
and hep-lat/9312054.
%
\bibitem{fleming}
T. Fleming, M. Gross, and R. Renken, Phys. Rev. D {\bf 50}, 7363 (1994).
%
\bibitem{homo}
W. Beirl, A. Hauke, P. Homolka, B. Krishnan, H. Kr\"oger, H. Markum,
and J. Riedler, Nucl. Phys. B (Proc. Suppl.) {\bf 47}, 625 (1996);
 W. Beirl, A. Hauke, P. Homolka, H. Markum, and J. Riedler,
Nucl. Phys. B (Proc. Suppl.) {\bf 53}, 735 (1997);
J. Riedler, W. Beirl, E. Bittner, A. Hauke, P. Homolka, and H. Markum,
Class. Quantum Grav. {\bf 16}, 1163 (1999).
%
\bibitem{bittnerzn}
E. Bittner, H. Markum, and J. Riedler, Nucl. Phys. B (Proc. Suppl.) {\bf 73}, 789
(1999).
%
\bibitem{gross}
M. Gross and H.W. Hamber, Nucl. Phys. B {\bf 364}, 703 (1991).
%
\bibitem{holm}
C. Holm and W. Janke, Phys. Lett. B {\bf 335}, 143 (1994).
%
\bibitem{physA}
E. Bittner, W. Janke, H. Markum, and J. Riedler, Physica A {\bf 277}, 
204 (2000).
%
\bibitem{ferrenberg}
A.M. Ferrenberg and R.H. Swendsen, Phys. Rev. Lett. {\bf 61}, 2635 (1988).
%
\bibitem{gaunt79}
D.S. Gaunt, M.F. Sykes, and S. McKenzie, J. Phys. A {\bf 12}, 871 (1979).
%
\bibitem{gaunt80}
S. McKenzie and D.S. Gaunt, J. Phys. A {\bf 13}, 1015 (1980).
%
\bibitem{velasco}
E. S\'anchez-Velasco, J. Phys. A {\bf 20}, 5033 (1987).
%
\bibitem{ralph}
R. Kenna and C.B. Lang, Phys. Lett. B {\bf 264}, 396 (1991);
Nucl. Phys. B {\bf 393}, 461 (1993).
%
\bibitem{parisi}
H.G. Ballesteros, L.A. Fern\'andez, V. Mart\'\i n-Mayor, A. Mu\~noz Sudupe, 
G. Parisi, and J.J. Ruiz-Lorenzo,
Nucl. Phys. B {\bf 512}, 681 (1998).
%
\bibitem{wolff}
U. Wolff, Phys. Rev. Lett. {\bf 62}, 361 (1989);
Nucl. Phys. B {\bf 322}, 759 (1989).
%
\bibitem{staufad}
D. Stauffer and J. Adler, Int. J. Mod. Phys. C {\bf 8}, 263 (1997).
%
\bibitem{gaunt}
D.S. Gaunt, M.F. Sykes, and S. McKenzie, J. Phys. A {\bf 12}, L339 (1976).
%
\bibitem{adstauf}
J. Adler and D. Stauffer, Int. J. Mod. Phys. C {\bf 6}, 807 (1995).
%
\bibitem{hamber00}
H.W. Hamber, Phys. Rev. D {\bf 61}, 124008 (2000).
%
\bibitem{berlin}
E. Bittner, W. Janke, and H. Markum,
Nucl. Phys. B (Proc. Suppl.) {\bf 106-107}, 989 (2002).
%
\end{thebibliography}
\end{document}